\documentclass[aps,pra,reprint,superscriptaddress,amsmath,amssymb]{revtex4-2}

\usepackage{graphicx}
\usepackage{listings}
\usepackage{longtable}
\usepackage{color,soul}
\usepackage{algorithm}
\usepackage[bookmarks,bookmarksnumbered,colorlinks,allcolors=blue]{hyperref}
\usepackage{bookmark}
\usepackage{makecell}

\definecolor{dkgreen}{rgb}{0,0.6,0}
\definecolor{gray}{rgb}{0.5,0.5,0.5}
\definecolor{mauve}{rgb}{0.58,0,0.82}

\begin{document}

\title{\Large Ultrahigh inductance superconducting materials from spinodal decomposition}

\author{Ran Gao}
\email{gaor410@gmail.com}
\affiliation{Alibaba Quantum Laboratory, Alibaba Group, Hangzhou, Zhejiang 311121, P.R.China}

\author{Hsiang-Sheng Ku}
\affiliation{Alibaba Quantum Laboratory, Alibaba Group, Hangzhou, Zhejiang 311121, P.R.China}

\author{Hao Deng}
\affiliation{Alibaba Quantum Laboratory, Alibaba Group, Hangzhou, Zhejiang 311121, P.R.China}

\author{Wenlong Yu}
\affiliation{Alibaba Quantum Laboratory, Alibaba Group, Hangzhou, Zhejiang 311121, P.R.China}

\author{Tian Xia}
\affiliation{Alibaba Quantum Laboratory, Alibaba Group, Hangzhou, Zhejiang 311121, P.R.China}

\author{Feng Wu}
\affiliation{Alibaba Quantum Laboratory, Alibaba Group, Hangzhou, Zhejiang 311121, P.R.China}

\author{Zhijun Song}
\affiliation{Alibaba Quantum Laboratory, Alibaba Group, Hangzhou, Zhejiang 311121, P.R.China}

\author{Xiaohe Miao}
\affiliation{Instrumentation and Service Center for Physical Sciences, Westlake University, Hangzhou, Zhejiang 310024, P.R.China}

\author{Chao Zhang}
\affiliation{Instrumentation and Service Center for Physical Sciences, Westlake University, Hangzhou, Zhejiang 310024, P.R.China}

\author{Yue Lin}
\affiliation{Hefei National Laboratory for Physical Sciences at the Microscale, University of Science and Technology of China, Hefei, Anhui 230026, P.R.China}

\author{Yaoyun Shi}
\affiliation{Alibaba Quantum Laboratory, Alibaba Group USA, Bellevue, Washington 98004, USA}

\author{Hui-Hai Zhao}
\affiliation{Alibaba Quantum Laboratory, Alibaba Group, Hangzhou, Zhejiang 311121, P.R.China}

\author{Chunqing Deng}
\email{dengchunqing@gmail.com}
\affiliation{Alibaba Quantum Laboratory, Alibaba Group, Hangzhou, Zhejiang 311121, P.R.China}

\begin{abstract}

\noindent Disordered superconducting nitrides with kinetic inductance have long been considered a leading material candidate for high-inductance quantum-circuit applications. Despite continuing efforts in reducing material dimensions to increase the kinetic inductance and the corresponding circuit impedance, it becomes a fundamental challenge to improve further without compromising material qualities. To this end, we propose a method to drastically increase the kinetic inductance of superconducting materials via spinodal decomposition while keeping a low microwave loss. We use epitaxial Ti\textsubscript{0.48}Al\textsubscript{0.52}N as a model system, and for the first time demonstrate the utilization of spinodal decomposition to trigger the insulator-to-superconductor transition with a drastically enhanced material disorder. The measured kinetic inductance has increased by 2-3 orders of magnitude compared with all the best reported disordered superconducting nitrides. Our work paves the way for substantially enhancing and deterministically controlling the inductance for advanced superconducting quantum circuits.

\end{abstract}

\maketitle
\bookmarksetup{startatroot}

\section{Introduction}

Superconducting materials with high kinetic inductance offer new opportunities in the design of high-sensitivity photon detectors, wideband quantum amplifiers, and high-coherence quantum processors~\cite{Manucharyan2009,Day2003,HoEom2012,Macklin2015}. In particular, for the noise-resilient superconducting-qubits, a non-dissipative circuit element as characterized by a considerably high inductance (\textit{i.e.,} several hundred nanohenry to a few microhenry) is indispensable for charge- and flux-noise protection~\cite{Earnest2018,Kalashnikov2020,Pechenezhskiy2020,Groszkowski2018,Gyenis2021a,Gyenis2021}. The magnitude of such inductance, as it is often the case, utterly exceeds the achievable geometric inductance in the nanofabricated superconducting circuits, making the utilization of intrinsic material properties such as kinetic inductance essential for its practical realizations. Extensive research have thus been performed on superconducting materials with a large kinetic inductance, such as Josephson-junction arrays (JJAs) \cite{Nguyen2019,Pop2014,Gyenis2021,Pechenezhskiy2020} and granular aluminum \cite{Grunhaupt2019,Kamenov2020,Moshe2020,Maleeva2018}, for the high-inductance implementations. 

In addition to aluminum-based materials, disordered superconducting nitrides (\textit{e.g.,} TiN, NbN, NbTiN) are also drawing increasing interests. Owing to the low energy dissipation, resilience to magnetic fields, and process compatibility, these emerging materials are advantageous for high-coherence applications and large-scale circuit integration~\cite{Samkharadze2016,Hazard2019,Peltonen2018,Shearrow2018,Knehr2021}. Nevertheless, due to the relatively small kinetic inductance (\textit{i.e.}, on the order of picohenry-per-square), a nanowire with large aspect ratio is inevitably required for the implementation of high-inductance elements~\cite{Chand2009,Coumou2013a,Hazra2018,Niepce2019,Pita-Vidal2020}. Any imperfection of the nanowire could be a source of energy dissipation and deteriorate the circuit coherence. As a result, tremendous research efforts have been made in the geometric control and optimization of these materials for the practical applications ~\cite{Samkharadze2016, Niepce2019, Yu2021, Pita-Vidal2020}.

A strategy to alleviate the stringent geometric constraints is to further increase the intrinsic kinetic inductance. Primarily, this was attained by either reducing material thickness or increasing the disorder. For instance, thin films of TiN or NbN have been shown to yield a sheet inductance of several hundred picohenry-per-square as the film is thinner than 10~nm~\cite{Shearrow2018,Coumou2013a,Coumou2013,Yu2021}. The chemical and structural disorder in NbTiN can also be increased if the deposition techniques and parameters are properly refined~\cite{Tsavdaris2017,Hazra2018}. However, despite the continuing optimization in the material synthesis, it remains a fundamental challenge to control the homogeneity of ultrathin films where a slight variation in thickness will inevitably result in substantial changes in the material properties. It is also known that ultrathin and disordered superconducting films are accompanied by large spacial fluctuations on the superconducting order parameter~\cite{Coumou2013,Kang2011,Carbillet2016,Carbillet2020}, which could trap quasiparticles in the superconductors and act as detrimental two-level systems~\cite{DeGraaf2020}. Moreover, the high defect density at the surfaces and edges of ultrathin nanowires typically plays a more significant role in deteriorating the microwave performance due to the larger filing factors of electric-field energy at these two regions~\cite{Gao2008,Samkharadze2016,Niepce2019,Coumou2013a,Niepce2020}. As such, aside from the augmentation of kinetic inductance by solely geometric reduction, an approach to intrinsically enhance the material disorder without compromising the material quality is desired.

\begin{figure*} [htp]
      \centering
      \includegraphics[width=12.9cm]{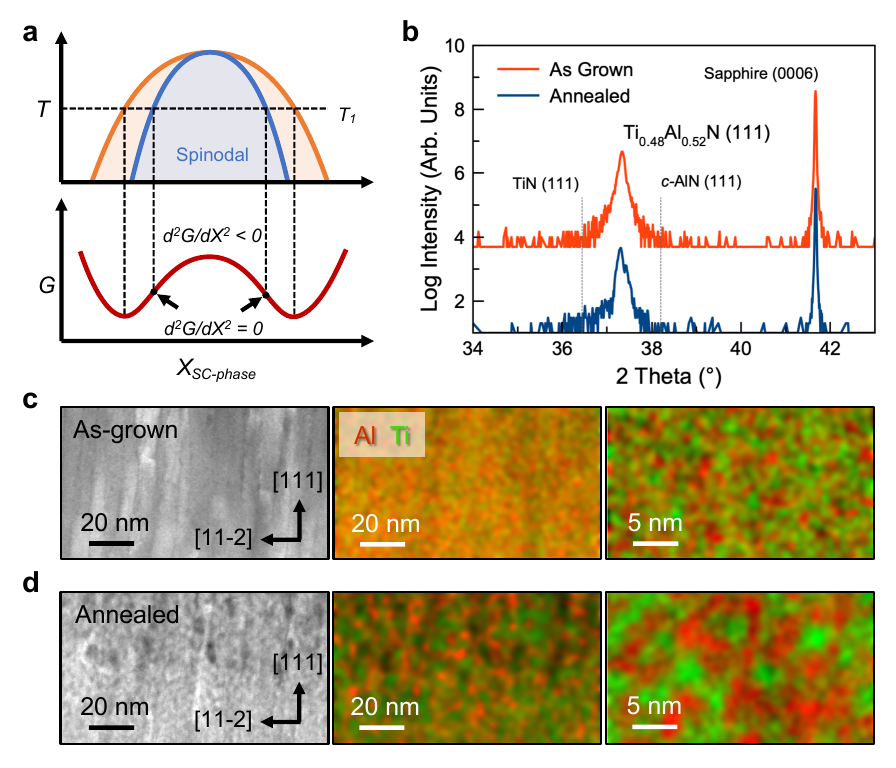}
      \caption{(a) The spinodal decomposition regime is illustrated in the miscibility gap of a phase diagram where the second derivative of \textit{Gibbs} free energy at a specific temperature (\textit{e.g.,} $T_1$) is negative. (b) X-ray line scans on 100nm-thick as-grown and annealed TAN films. (c) The HAADF-STEM and STEM-EDS scans for as-grown TAN films. (d) The HAADF-STEM and STEM-EDS scans for annealed TAN films. Clear phase segregation was revealed by HAADF-contrast variations and STEM-EDS mapping.}
      \label{fig:TANMaterFig1} 
\end{figure*}

In this work, we demonstrate a method to substantially increase the disorder and the kinetic inductance of epitaxial-quality superconducting materials with low microwave losses. The idea is to utilize the spinodal decomposition in a compound where the spontaneous phase segregation results in a much stronger chemical and structural disorder. An extensively studied material Ti\textsubscript{0.48}Al\textsubscript{0.52}N (TAN) was used as a model system to verify the feasibility of this approach. Although this material has long been considered for hard coating~\cite{Zhou2017,ToBaben2017}, the combination of TiN (which is a superconductor) and AlN (which is a wide band-gap insulator) makes it ideal for the introduction of low-loss non-superconducting phases in the superconducting phases. We show that, the as-grown TAN epitaxial films on sapphire substrates were insulating, while after the phase segregation, an insulator-to-superconductor transition was observed. Structural characterization on the phase-segregated materials revealed good crystallinity and a random distribution of Ti-rich and Al-rich regions, consistent with our simulations and the previous studies on similar spinodal systems~\cite{Knutsson2013,Zhou2017}. Subsequent transport and cryogenic measurements on microwave resonators have shown a sharply enhanced sheet inductance of $L_k^{\square} \approx57$ pH/$\square$ for the TAN films as thick as 100 nm, and a reasonably good resonator quality factor of $Q_i \sim2.7\times10^5$. As the film thickness was reduced to intermediate value of 20 nm, the sheet inductance was measured to be as large as $\sim7.5$ nH/$\square$, nearly 2-3 orders of magnitude higher than all the reported nitride superconductors and comparable to the highest sheet inductance ever achieved in the granular-aluminum system~\cite{Shearrow2018,Coumou2013a,Niepce2019,Hazard2019,Samkharadze2016,Hazra2018,Maleeva2018,Kamenov2020,Moshe2020,Zhang2019,Valenti2019}. Our results demonstrate that the spinodal TAN films are promising candidates for the implementation of high-inductance and low-dissipation circuit elements. With a systematic engineering over the chemical compositions and micro-structures, spinodal superconducting systems may offer an unprecedented opportunity in the high-impedance superconducting quantum circuits.

\section{Spinodal decomposition and structural characterization}

Spinodal decomposition is a spontaneous phase segregation process in mixtures where an infinitesimal perturbation can drive the segregation of phases without nucleation. Given a homogeneous solid mixture composed of two thermodynamically stable phases, the spinodal decomposition takes place in a range of compositions where the second derivative of the total Gibbs free energy is negative (Fig.~\ref{fig:TANMaterFig1}a). Within this regime, any infinitesimal change in the material composition reduces the total energy, giving rise to a random distribution of segregated phases. Such phase segregation, in turn, creates a stronger chemical and structural disorder in the system and drastically changes the material properties. For example, given that one of the phases is a superconductor (\textit{e.g.,} TiN) and the other is an insulator (\textit{e.g.,} AlN), we have simulated the redistribution of atoms and acquired a microscopic structure with the Ti-rich regions largely confined by Al-rich regions (Supplementary Fig.~1). It can be further hypothesized that, as the phase segregation evolves, an insulator-to-superconductor transition could be observed in the Ti-rich regions, and the spinodal system manifests as a highly-disordered superconducting material. In this spirit, we carried out a set of detailed studies on a model epitaxial TAN films to verify this hypothesis.

100nm-thick epitaxial TAN thin films with a nominal stoichiometry Ti\textsubscript{0.5}Al\textsubscript{0.5}N were grown on the $c$-cut sapphire substrates using a magnetron co-sputtering system (see Methods). Part of the deposited wafers from the same batch were annealed for phase segregation as comparative studies (\textit{i.e.,} 1000 $^{\circ}$C for 30 mins, see Methods). From the X-ray line scans (Fig.~\ref{fig:TANMaterFig1}b, Supplementary Fig.~2a), the as-grown TAN films are revealed to be epitaxial and single-phase as indicated by the clean and sharp diffraction peaks at the 111-diffraction conditions. Based on the diffraction peak positions of the two parent phases (\textit{i.e.,} TiN and \textit{cubic}-AlN~\cite{Xia1998,Camp1991,Gao2021}), the nominal stoichiometry of TAN films is achieved according to Vegard's law, and was quantitatively confirmed to be Ti\textsubscript{0.48}Al\textsubscript{0.52}N by X-ray photo-electron spectroscopy (XPS) depth-profile studies (Supplementary Fig.~3). For the annealed TAN samples, the diffraction peaks remained almost unchanged, suggesting that the crystallinity of the samples is maintained even after the phase segregation. The good quality of the as-grown and annealed films were additionally revealed by rocking-curve studies, both yielding a full-width-half-max of $\sim$0.04$^{\circ}$ (Supplementary Fig.~ 2b). It is worth noting that, although a number of annealing conditions have been tested and will inevitably result in different level of phase segregation~\cite{Hoglund2010,Zhou2017}, a fixed annealing condition was carried out throughout this work for in-depth material investigations.

To better understand the microstructure evolution, high-angle annular dark-field scanning transmission electron microscopy (HAADF-STEM) was performed on both as-grown and annealed samples. The HAADF images along $[01\bar{1}]_{\text{TAN}}$ zone axis (Fig.~\ref{fig:TANMaterFig1}c) revealed clear columnar nature of the as-grown TAN films, where the columnar domains exhibited good crystallinity as indicated by the sharp fast-\textit{Fourier}-transform patterns and the clearly visible atomic columns (Supplementary Fig.~2c). Note that the columnar domains are mainly tilted rotational domains offset from each other by 60$^{\circ}$ along the out-of-plane $[111]_{\text{TAN}}$-axis, which have been studied in detail for epitaxial TiN/sapphire heterostructures~\cite{Gao2021}. After the thermal annealing (Fig.~\ref{fig:TANMaterFig1}d), the macroscopic domain structures remained essentially unchanged while the contrasts of the HAADF images are turning inhomogeneous within each domain. The variation in HAADF contrasts is a result of the phase segregation where regions with distinct atomic masses scatter the electron beam differently, and the dislocation and micro-strain evolution accompanied with the atomic diffusion processes~\cite{Calamba2019,Calamba2021,Chaar2021}. To get a better picture of the atomic distribution, energy dispersive spectroscopy (STEM-EDS) on the atomic ratio of titanium and aluminum was performed. It can be shown from the STEM-EDS mapping that the as-grown TAN films possess a relatively homogeneous distribution of titanium and aluminum atoms (Fig.~\ref{fig:TANMaterFig1}c). After the annealing (Fig.~\ref{fig:TANMaterFig1}d), more distinct Ti-rich and Al-rich regions with sizes on the order of nanometers can be observed. Besides the microscopic phase segregation, a gentle gradient of elemental distribution can be observed from the film-substrate interface to the film surface of annealed samples (Fig.~\ref{fig:TANMaterFig1}d, Supplementary Fig.~3b). This could be a result of various convoluted effects such as changes in diffusion energy barriers due to the variation in the micro-strains from the interface to the surface~\cite{Calamba2019a,Calamba2019} and the difference in affinity to oxygen between aluminum and titanium atoms~\cite{Greczynski2019}. As the concentration gradient is reasonably small (less than 5\%), its impacts are neglected in this work and subject to future studies. In all, high-quality epitaxial TAN thin films have been synthesized as a model spinodal system, and clear phase segregation was observed upon thermal annealing without degrading the structural qualities. With a detailed understanding over the structural and chemical properties, we proceeded to examine the electrical behaviors of the material.

\section{Electrical transport studies}

\begin{figure*} [t]
      \centering
      \includegraphics[width=12.9cm]{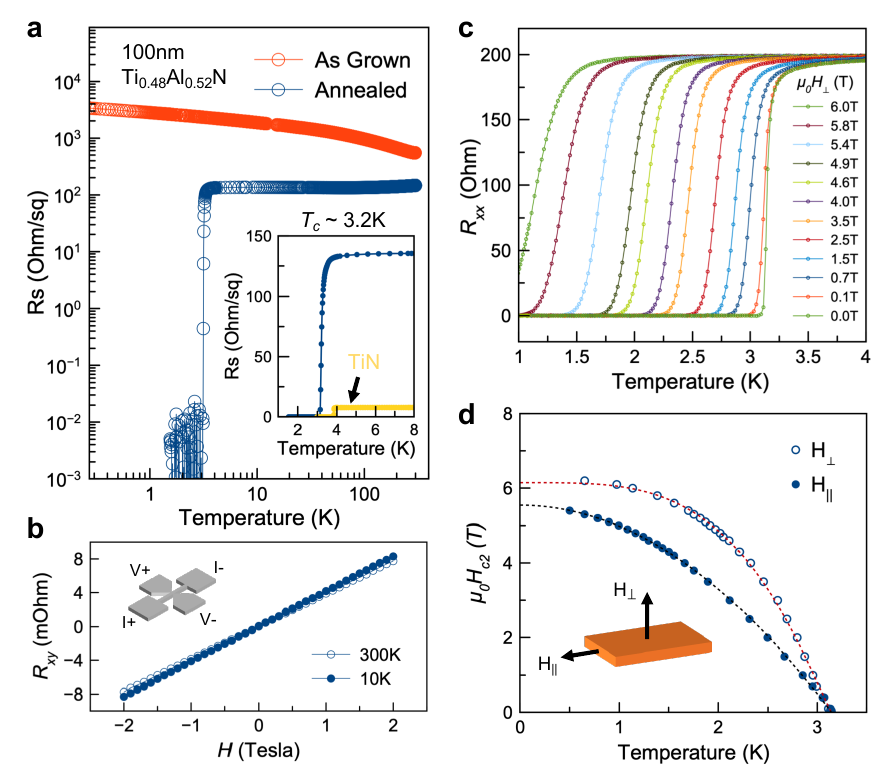}
      \caption{(a) The temperature-dependent sheet resistance $R_s(T)$ for both as-grown and annealed 100 nm-thick TAN films. An insulator-to-superconductor transition was introduced by phase segregation with a transition temperature $T_c$=3.2 K (inset, where the $R_s(T)$ of 100 nm-thick TiN parent phase is given for comparison). (b) \textit{Hall} resistance as a function of applied magnetic field for annealed TAN films. A clean and positive slope with small temperature dependency revealed the metallic nature of annealed samples. (c) Temperature-dependent longitudinal resistance ($R_{xx}(T)$) under varied magnetic field applied perpendicular to the substrate surface. (d) The relationships between the upper critical ($H_{c2}$) and superconducting transition temperature ($T_c$) are given for both perpendicular and parallel fields. Dotted lines are the fits to the experimental data.}
      \label{fig:TANMaterFig2} 
\end{figure*}

The electrical measurements were performed on samples patterned with \textit{Hall}-bar geometry (see Methods, Supplementary Fig.~4). The temperature-dependent $R_s(T)$ curves from 300~K to 200~mK were first acquired for both as-grown and annealed TAN films (Fig.~\ref{fig:TANMaterFig2}a). The as-grown TAN samples showed a negative $dR/dT$, indicating the insulating nature of the homogeneous Ti-Al-N mixture. After the annealing, an insulator-to-superconductor transition was observed with a relatively sharp phase transition temperature of $T_c\approx$ 3.2~K and a sheet resistance of $R_s\approx$ 132~$\Omega/\square$. Comparing with the 100 nm-thick TiN parent phase deposited under the same conditions, there is a 2 orders of magnitude increase in the sheet resistance (Fig.~\ref{fig:TANMaterFig2}a, inset). Remarkably, this observation has confirmed our hypothesis that the creation of localized Ti-rich regions in TAN spinodal systems can indeed give rise to superconductivity with significantly enhanced disorder. As a result, the sheet kinetic inductance of this 100 nm-thick TAN film given by $L_k^\square=\frac{\hbar R_s}{\pi\Delta}$ is expected to be as large as 57 pH/$\square$, a value comparable to those observed in much thinner nitride superconductors. 

A careful analysis on the temperature dependence of conductivity for the annealed TAN films revealed the material to be a disordered metal (see Supplementary Materials, Supplementary Fig.~5)~\cite{Kaveh1982,Lee1985}. This is followed and confirmed by \textit{Hall} resistance measurement (Fig.~\ref{fig:TANMaterFig2}b) that yields a nearly temperature-independent carrier concentration of $n_e^{10K}=1.6\times 10^{28}$ m$^{-3}$ and an elastic mean-free-path of $l$ = 0.18~nm (corresponding to a $k_Fl$ = 1.38) under the assumption of the free-electron model (see Supplementary for more detailed discussion). To further examine the superconducting properties of the annealed TAN films, longitudinal resistance $R_{xx}(T)$ was measured under a series of magnetic fields applied both in parallel and perpendicular to the substrate surface (Fig.~\ref{fig:TANMaterFig2}c for perpendicular field and Supplementary Fig.~6 for parallel field). The superconducting transition temperature was chosen where the longitudinal resistance was reduced by half, and the relationships between the upper critical fields and the superconducting transition temperatures ($H_{c2}^{\parallel,\perp}(T)$) were plotted (Fig.~\ref{fig:TANMaterFig2}d). For the parallel fields, $H_{c2}^{\parallel}(T)$ curve can be well fitted by the \textit{Ginzburg-Landau} (\textit{GL}) formula $\mu_0H_{c2}^{\parallel}(T)=\Phi_0/2\pi\xi_{GL}^2[1-(T/T_c)^2]$ and the \textit{GL coherence length} was extracted to be $\xi_{GL}$ = 7.7 nm. Owing to the fact that $l\ll\xi_{GL}\ll\lambda_L$, where the \textit{London penetration depth} $\lambda_L$ for disordered superconductors is typically hundreds of nanometers~\cite{Qiao2018,Diener2012,Pracht2012}, the material is confirmed to be a bulk-like, dirty type-II superconductor. While for the perpendicular fields, a deviation from the empirical \textit{GL} formula was observed where the curve has an upward curvature at low fields and saturates to a higher critical fields with weaker temperature dependency (fitting details see Supplementary materials, Supplementary Fig.~6). We propose that the slight anisotropy between the parallel and perpendicular fields is caused by the existence of vertical columnar grain boundaries (see Fig.~\ref{fig:TANMaterFig1}c,d and Supplementary Fig.~2c,d). Namely, at low fields, the columnar structures make the material resemble those granular and weakly-coupled superconductors, featuring an upward curvature in $H_{c2}^{\perp}(T)$ due to the change of relative magnitudes between the inter-grain coupling energy and the magnetic-flux energy~\cite{Deutscher1978,Deutscher1980,Klemm1975}. As the field increases, the grain boundaries serve as pinning centers for the aligned vortices and enhance the superconductivity robustness~\cite{Jung2012,Matsumoto2009,Durrell2011}. To conclude, the phase segregation induced by spinodal decomposition in TAN results in a insulator-to-superconductor transition at around 3.2~K and a largely enhanced kinetic inductance of the material.

\section{Cryogenic microwave resonator studies}

\begin{figure} [t]
      \centering
      \includegraphics[width=8.6cm]{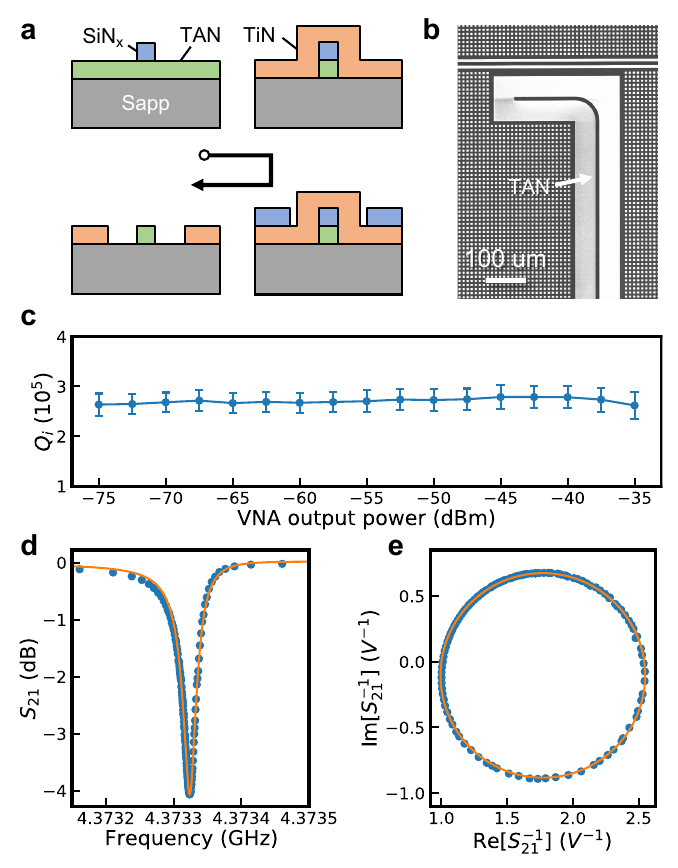}
      \caption{(a) The schematic illustration of the integration of TAN into the established TiN processes with silicon nitride hardmask and and wet etchants. (b) Scanning electron microscopy image of the fabricated microwave CPW resonator. The readout feed line and ground plane were made with low-loss epitaxial TiN films while the high-inductance center resonator line was made with 100 nm-thick annealed TAN. (c) Ten TAN resonators showed an averaged quality factor $Q_i$ of $\sim2.7\times10^5$ with little photon-number dependency, and (d-e) the exemplary fittings for $Q_i$ extraction of one of the resonators was given.}
      \label{fig:TANMaterFig3} 
\end{figure}

To examine the applicability of TAN for superconducting quantum circuits, we first demonstrate that the TAN system can be easily integrated into our established TiN-based processes~\cite{Gao2021}. Briefly, the TAN films were wet etched in a solution of ammonia hydroxide and hydroperoxide masked by a layer of SiN\textsubscript{x}. The etched wafers were then sent for TiN deposition without hardmask stripping. The deposited TiN layer was subsequently patterned using the SiN\textsubscript{x}-hardmask and the wet etchants (Fig.~\ref{fig:TANMaterFig3}a, details see Methods). The eventual device structure was obtained after rinsing off the hardmask in the DHF solutions, and the scanning electron microscopy images of one exemplary device was illustrated (Fig.~\ref{fig:TANMaterFig3}b). As given in the figures, the coplanar waveguide (CPW) resonators features a center resonator line (10 $\mu$m in width, 50 $\mu$m gap to ground) made of 100 nm-thick TAN films and a ground plane together with the readout feed line made of 100 nm-thick TiN films. 

Cryogenic microwave studies on the CPW resonator devices were performed using a vector-network analyzer (cryogenic measurement setup see Supplementary Fig.~7). The transmission scattering parameter $S_{21}$ of the feed line was collected and the internal quality factor $Q_i$ of the capacitively coupled TAN resonators was extracted by fitting the inverse $S_{21}$ spectra given by $S_{21}^{-1}(f) = 1 + \frac{Q_i}{Q_c^*}e^{i\phi} \frac{1}{1+2 i Q_i \frac{f-f_0}{f_0}}$, where $Q_i = (Q^{-1}-Q_c^{-1})^{-1}$, $Q_c^*=Q_c (Z_0/|Z|)$, $Z_0$ is the cable impedance, and $Z = |Z|e^{i\phi}$ is half the inverse sum of the environment impedance on the input and output sides of the resonators. Averaged quality factors of multiple resonators as a function of calculated photon numbers in the resonators were plotted (Fig.~\ref{fig:TANMaterFig3}c), and the fitting spectra of one exemplary resonator were also provided (Fig.~\ref{fig:TANMaterFig3}d,e). The averaged quality factor at single-photon limit is extracted to be $Q_i\approx2.7\times10^5$, comparable and even better other high-kinetic-inductance resonators~\cite{Samkharadze2016,Zhang2019,Niepce2019,Maleeva2018}. In addition, the resonator quality factors were observed to have little photon-number dependency, indicating that an energy relaxation mechanism other than the well-known two-level systems (TLS) could be dominant and is under investigation for future studies. Moreover, by comparing the measured resonator frequencies with the circuit-electromagnetic simulation using HFSS (\textit{ANSYS}, see Methods), it was found that the extracted surface inductance of TAN resonator is 57.4 pH/$\square$, matching well with that predicted in the transport studies. Altogether, a combination of electrical transport studies and cryogenic microwave resonator studies have all revealed the spinodal TAN system as a promising material candidate to be used for high-inductance and low-microwave-loss circuits. 

\section{Discussions and conclusions}

\begin{figure} [t]
      \includegraphics[width=8.6cm]{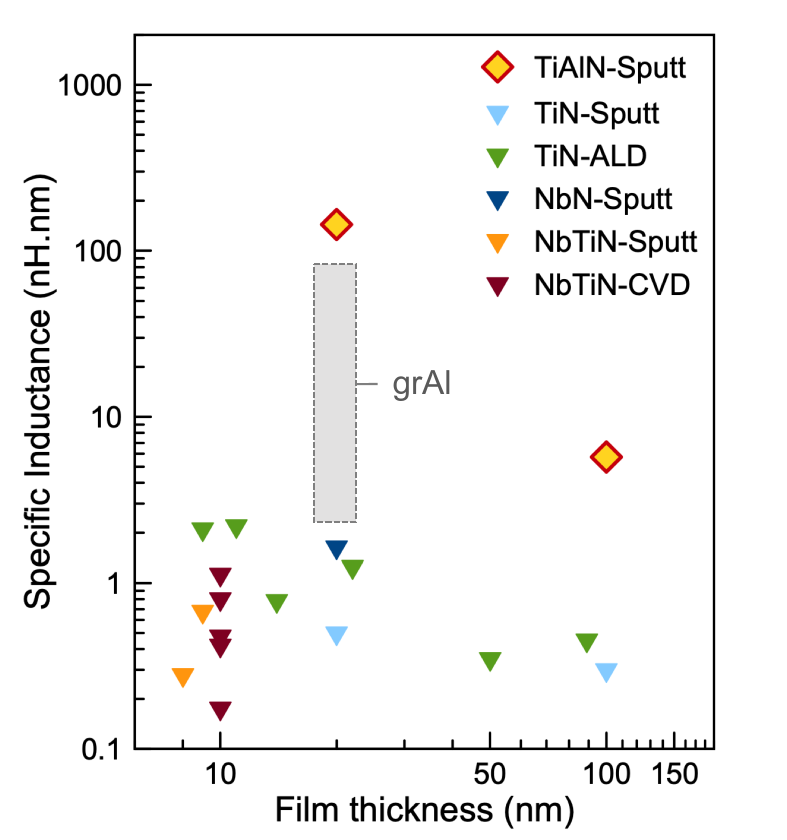}
      \caption{The specific inductance and the corresponding film thickness of various materials including TAN (this work), TiN~\cite{Shearrow2018,Coumou2013a,Gao2021}, NbTiN~\cite{Samkharadze2016,Hazard2019,Hazra2018}, NbN~\cite{Niepce2019}, and granular aluminum~\cite{Maleeva2018,Kamenov2020,Moshe2020,Zhang2019}. The granular aluminum has a wide range of reported values and is indicated by the shaded area.}
      \label{fig:TANMaterFig4} 
\end{figure}

We further compare the TAN system with other reported high-kinetic-inductance material systems including a number of disordered superconducting nitrides~\cite{Shearrow2018,Coumou2013a,Niepce2019,Hazard2019,Samkharadze2016,Hazra2018} and granular aluminum~\cite{Maleeva2018,Kamenov2020,Moshe2020,Zhang2019,Valenti2019} (Fig.~\ref{fig:TANMaterFig4}). Since it is very often the case that only sheet inductance is reported in the literatures, making it burdensome to compare materials with varied thicknesses, we use specific inductance (\textit{i.e.,} sheet inductance $L_s$ times material thickness $t$) to better reflect intrinsic material properties. For strongly disordered thin films with large magnetic-field penetration depth ($\lambda \gg t$), the specific inductance given by $L_s\cdot t=\frac{\mu_0\lambda t}{2}\coth(\frac{t}{2\lambda})\approx \mu_0\lambda^2$ is essentially equivalent to the penetration depth of the materials. In the meantime, due to the fact that the penetration depth also scales with the film thickness, the specific inductance with the corresponding film thicknesses are indicated in the same plot. 

For all the superconducting nitrides grown via a number of deposition techniques, the specific inductance is no greater than $\sim$3 nH$\cdot$nm. Although different level of chemical and structural disorder could be induced in these material systems by varying the chemical compositions and synthesis approaches, there appears to be a fundamental limit of the achievable specific inductance without the degradation of material qualities. While for spinodal TAN system with high degree of crystallinity, the 100 nm-thick films have a specific inductance of 5.7 nH$\cdot$nm, already higher than all the other reported ultrathin nitride materials. Again, although the specific inductance of TAN is significantly enhanced, the microwave losses as probed by CPW resonators remain low and comparable to other high-kinetic-inductance materials with much smaller specific inductance. To further examine thinner films and the potential of this material system, 20 nm-thick TAN films with the same stoichiometry were measured (Supplementary Fig.~8) and revealed a specific inductance as large as $\sim$144 nH$\cdot$nm (corresponding to a sheet inductance of 7.5 nH/$\square$). This value is nearly 3 orders of magnitude higher compared with the parent 20 nm-thick TiN phase and larger than many of the reported granular-aluminum systems. Given that only one representative combination of material stoichiometry and annealing condition was used in this study, it is reasonable to expect that this work has yet set any limit on the specific inductance of spinodal materials. As such, with a comprehensive engineering over the chemical compositions and phase segregation, we believe that the specific inductance of such superconducting spinodal materials could be readily tuned and controlled over a tremendously wide range.

To conclude, we have demonstrated that spinodal phase segregation in a model Ti\textsubscript{0.48}Al\textsubscript{0.52}N system can give rise to an insulator-to-superconductor transition and a drastically increased material disorder compared with the TiN parent system. The chemical and structural characterization reveal that the material remain epitaxially crystalline, while a sharp increase in the specific inductance (\textit{i.e.,} 2-3 orders of magnitudes larger compared with other disordered superconducting nitrides) is confirmed by both electrical transport studies and cryogenic microwave-resonator studies. To the best of our knowledge, this is the first demonstration of utilizing spinodal decomposition to trigger the insulator-to-superconductor transition with an enhanced kinetic inductance and low microwave losses. The ultrahigh kinetic inductance and good process compatibility of spinodal TAN mitigates the challenges in the engineering of low-dimension structures, making it a suitable material candidate for the development of noise-resilient superconducting qubits and high-impedance circuits. In addition, much as the large enhancement in kinetic inductance has been demonstrated in the model TAN system, similar engineering and optimization routes are in principle applicable to other spinodal systems for the investigation and optimization on materials and physical properties. This methodology, in turn, offers an unprecedented opportunity to deterministically design and engineer low-loss and high-inductance materials for advanced superconducting-quantum-circuit applications.

\begin{acknowledgments}
We thank all members at Alibaba Quantum Laboratory for the fruitful discussions and supports. We thank the Westlake Center for Micro/Nano Fabrication and the Instrumentation and Service Center for Physical Sciences at Westlake University for the fabrication supports and the characterization services. We thank Dr. Xin Wan at Zhejiang University and Dr. Yanwei Cao at Ningbo Institute of Industrial Technology for the insightful discussions.
\end{acknowledgments}

\clearpage
\section{Methods}

\textbf{Thin-film deposition.} The sputter system used for this study is a CS-200 R\&D type (\textit{ULVAC Inc.}). The deposition chamber is equipped with a turbo-molecular pump that reaches a base-pressure of $5 \times 10^{-5}$ Pa. The deposition of TAN thin films was performed using a co-sputtering approach with a RF sputter gun using a 99.999\% titanium target and a DC sputter gun using a 99.999\% aluminum target. Before the deposition, the as-received Sapphire substrates (2-inch, single-side polished, \textit{Suzhou RDMICRO Co., Ltd.}) were cleaned and heated to 300 $^\circ$C in vacuum for dehydration. The sputter guns were then pre-sputtered with shutter closed for 60 sec to remove surface contaminants. After deposition, the samples were cooled for 10 mins and unloaded from the load-lock by venting with nitrogen. 

\textbf{Annealing setup.} The thermal annealing was performed in a tube furnace equipped with a alumina tube and a vacuum pump. Samples were mounted in the center of the hot zone and the tube was then sealed and purged with argon (5N-purity) for at least three times. A constant flow of argon was maintained throughout the anneal at atmospheric pressure. The annealing conditions were fixed at 1000 $^\circ$C for 30 mins. 

\textbf{X-ray diffractometry (XRD).} The XRD studies including linescans and rocking curves were performed on a high-resolution D8 ADVANCE X-ray diffraction system (\textit{Bruker}) with optics set up for epitaxial-film studies. 

\textbf{Transmission electron microscopy (TEM).} Cross section samples were prepared with the focused ion beam (FIB) technique on an FEI Helios 650 system. The high-angle annular dark-field (HAADF) images were acquired using JEOL JEM-ARM200F electron microscope, which is equipped with a thermal field-emission gun and spherical aberration (Cs) corrector. The semi-convergence angle is around 23 mrad during HAADF imaging, and the collection angles of the detectors are 50 and 200 mrad. Energy-dispersive X-ray spectroscopy (EDX) mapping was carried out on an FEI Talos F200X TEM at 200kV equipped with a 4-inch column SDD Super-X detectors. 

\textbf{X-ray photo-electron spectroscopy (XPS).} The XPS studies were performed on a ESCALAB Xi+ XPS system (\textit{Thermal Fisher}) at an incident angle of 60 $^{\circ}$. The system was first calibrated with standard samples, and the selective milling between titanium, aluminum, oxygen, and nitrogen was confirmed to be minimal. For the XPS depth-profile studies, the milling area was set to be 2 mm by 2 mm while the analyzing area was concentric with the milling area and set to be 0.4 mm by 0.4 mm. High milling current of the $Ar^+$ was applied, and the beam energy was set to 2000 eV. The spectra were taken after every milling step in a 20 sec interval for the surface region and 300 sec interval for the bulk part.  

\textbf{Device fabrication.} The devices for transport studies were patterned into \textit{Hall}-bar geometries using an established hardmask-based wet-etch approach (Supplementary Fig.~4)~\cite{Gao2021}. First, the as-grown TAN films were coated with a layer of 100 nm-thick SiN\textsubscript{x} using a Haasrode-C200A PECVD system (\textit{LEUVEN Instruments}). The bi-layer stack, with lithographically patterned photoresist, was then loaded in a PlasmaPro100 ICP system (\textit{Oxford Instruments}) for SiN\textsubscript{x} patterning. After the hardmask patterning and photoresist stripping, the TAN films were then wet-etched by rinsing the masked-stack in the SC-1 solution heated to 60 $^{\circ}$C, which was then followed by hardmask stripping in the diluted hydrofluoric solutions (DHF). The yielded wafers were then coated with another layer of SiN\textsubscript{x} for protection over oxidation and annealed. The protective SiN\textsubscript{x} capping was stripped again in the DHF and the wafers were diced for further measurements. 

For the microwave-resonators with TAN films integrated into the TiN-based quantum circuits, it is essentially a repetition of the processes discussed above. First, the TAN films was wet-etch in the SC-1 solution without stripping off the SiN\textsubscript{x} hardmask. The wet-etched wafer was subsequently rinsed in distilled water, nitrogen-blow dried, and loaded into the sputter chamber for the epitaxial-TiN deposition. The TiN deposition was performed using a 99.999\% titanium target with RF sputter gun using established processes. After TiN deposition, the wafers were coated with a layer of SiN\textsubscript{x} hardmask, and was patterned similarly using the processes described above. The TiN layer was also wet-etched in the SC-1 solution, followed by hardmask stripping in the DHF solutions.

\textbf{Transport studies.} The transport studies were performed on patterned \textit{Hall}-bar samples and the electrical connections to the sample puck were made by aluminum wedge bonding (see Fig.~\ref{fig:TANMaterFig2}b, inset). The samples were then loaded in a physical-properties measurement system (\textit{Quantum Design Inc.}) equipped with a tilting sample stage for all the subsequent characterizations. A constant current of 10 $\mu$A was supplied for both longitudinal and \textit{Hall} resistance measurements.

\textbf{Cryogenic measurement setup.} A conventional cryogenic measurement setup for the microwave resonators studies were used (Supplementary Fig.~7). The microwave signal was generated by a E5071C ENA Vector Network Analyzer (VNA, \textit{Keysight Technologies}) and sent to the device under test (DUT) at the base temperature ($\sim10$ mK) through a series of attenuators and filters. The sample was isolated from the environment by a layer of Cu-Al shield and a layer of magnetic shield. The signal from the DUT was amplified using a high-electron-mobility transistor amplifier mounted at the 4 K-stage, and sent back to the VNA after an additional amplification at room temperature. 

\textbf{Electromagnetic simulation.} The circuit impedance and the corresponding kinetic inductance of the superconducting materials were extracted via electromagnetic simulation using HFSS (\textit{ANSYS}). The circuit model was set up based on the exact geometry of the resonator devices. By adding a continuously varied surface inductance on the resonator center line (and a known and fixed surface inductance on the TiN ground), the resonator frequencies shift towards the measured resonance frequencies until the surface inductance matches with the realistic values. The resulting surface inductance was taken as the sheet kinetic inductance of the TAN films.

\bibliography{TANMater_Ref}

\end{document}


\vspace {1cm}

\title{\LARGE Supplementary Materials \\
\Large Ultrahigh inductance materials from spinodal decomposition}

\author{Ran Gao}
\email{gaor410@gmail.com}
\affiliation{Alibaba Quantum Laboratory, Alibaba Group, Hangzhou, Zhejiang 311121, P.R.China}

\author{Hsiang-Sheng Ku}
\affiliation{Alibaba Quantum Laboratory, Alibaba Group, Hangzhou, Zhejiang 311121, P.R.China}

\author{Hao Deng}
\affiliation{Alibaba Quantum Laboratory, Alibaba Group, Hangzhou, Zhejiang 311121, P.R.China}

\author{Wenlong Yu}
\affiliation{Alibaba Quantum Laboratory, Alibaba Group, Hangzhou, Zhejiang 311121, P.R.China}

\author{Tian Xia}
\affiliation{Alibaba Quantum Laboratory, Alibaba Group, Hangzhou, Zhejiang 311121, P.R.China}

\author{Feng Wu}
\affiliation{Alibaba Quantum Laboratory, Alibaba Group, Hangzhou, Zhejiang 311121, P.R.China}

\author{Zhijun Song}
\affiliation{Alibaba Quantum Laboratory, Alibaba Group, Hangzhou, Zhejiang 311121, P.R.China}

\author{Xiaohe Miao}
\affiliation{Instrumentation and Service Center for Physical Sciences, Westlake University, Hangzhou, Zhejiang 310024, P.R.China}

\author{Chao Zhang}
\affiliation{Instrumentation and Service Center for Physical Sciences, Westlake University, Hangzhou, Zhejiang 310024, P.R.China}

\author{Yue Lin}
\affiliation{Hefei National Laboratory for Physical Sciences at the Microscale, University of Science and Technology of China, Hefei, Anhui 230026, P.R.China}

\author{Yaoyun Shi}
\affiliation{Alibaba Quantum Laboratory, Alibaba Group USA, Bellevue, Washington 98004, USA}

\author{Hui-Hai Zhao}
\affiliation{Alibaba Quantum Laboratory, Alibaba Group, Hangzhou, Zhejiang 311121, P.R.China}

\author{Chunqing Deng}
\email{dengchunqing@gmail.com}
\affiliation{Alibaba Quantum Laboratory, Alibaba Group, Hangzhou, Zhejiang 311121, P.R.China}

\maketitle

\section{Numerical simulation of spinodal decomposition process}

In this section, we perform numerical simulation of a simplified model to qualitatively explain the spinodal decomposition process observed in our experiment.
A continuous phase field approach in two dimensions is used where Ti\textsubscript{x}Al\textsubscript{1-x}N (TAN) is approximated as a pseudo-binary system consisting of TiN and AlN.
The microscopic structure evolution of the spinodal decomposition in TAN can be simulated by solving the modified Cahn–Hilliard equation\cite{Cahn1958free}
\begin{equation}
      \frac{\partial x(\vec{r },t)}{\partial t}
      = \nabla \left( D \nabla\left( \frac{\partial G(x)}{\partial x(\vec{r },t)} 
      - 2 \kappa \nabla^2 x(\vec{r}, t) \right)\right),
      \label{eq:Cahn–Hilliard}
\end{equation}
where $x(\vec{r})$ is the fraction of Ti at the position $\vec{r}$ of the TAN film, $G(x)$ is the Gibbs energy per unit area of homogeneous film as a function of $x$, $D$ is the diffusion coefficient, and $\kappa$ is the gradient energy coefficient.

\begin{figure} [h]
      \centering
      \includegraphics[width=12.9cm]{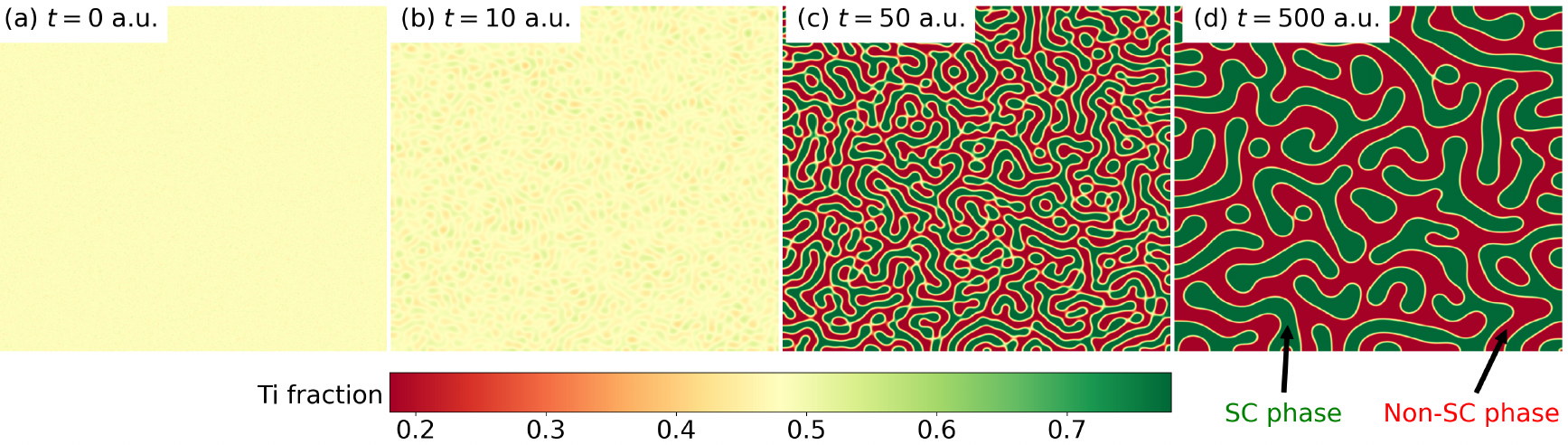}
      \caption{
      Simulated microscopic structure evolution of the spinodal decomposition process starting from a nearly homogeneous Ti\textsubscript{0.48}Al\textsubscript{0.52}N film, with red corresponding to Al and green to Ti. Figures (a), (b), (c) and (d) show the snapshots of the Ti fractions at evolution time steps $t=0$ a.u., $t=10$ a.u., $t=50$ a.u. and $t=500$ a.u. respectively.}
      \label{fig:TANMaterFigS1} 
\end{figure}

Supplementary Fig.~\ref{fig:TANMaterFigS1} shows 4 snapshots of the simulated microscopic structure evolution of the spinodal decomposition process starting from a nearly homogeneous Ti\textsubscript{0.48}Al\textsubscript{0.52}N film, with red corresponding to Al and green to Ti. 
In our simulation, we assume $D=\kappa=1$ a.u..
The initial configuration $x(\vec{r}, 0)$ is assumed to be nearly homogeneous with small fluctuation to mimic the stoichiometry variations in the epitaxial TAN thin films. We assume the initial configuration of Al fraction obeys Gaussian distribution, i.e. $x(\vec{r}, 0) \sim \mathcal{N}(\mu, \sigma^2)$, where we set the average value $\mu = 0.48$ according to our experiment, and assume small fluctuations with the variance $\sigma^2 = 10^{-3}$. To make the spinodal decomposition spontaneously occur, $d^2 G(x) / dx^2$ should be negative in a range of compositions. In our simulation, we assume the Gibbs energy has a phenomenological double-well form $G(x)=x^2(1-x)^2$, which is a very simply approximation but qualitatively agrees with that obtained from the first principle simulation of TAN crystal~\cite{Alling2011unified}. 
To solve the initial value problem of Supplementary Eq.~\ref{eq:Cahn–Hilliard} numerically, we apply the finite volume method and divide the spatial domain by uniform mesh scheme into $1000 \times 1000$ squares.
During the evolution, the film gradually decomposes into segregated Ti-rich domains confined by Al-rich domains, and the domains coarsening is observed, which qualitatively agrees with our experiment. 

\newpage
\section{Additional characterization data on TAN thin films}

\begin{figure} [H]
      \centering
      \includegraphics[width=12.9cm]{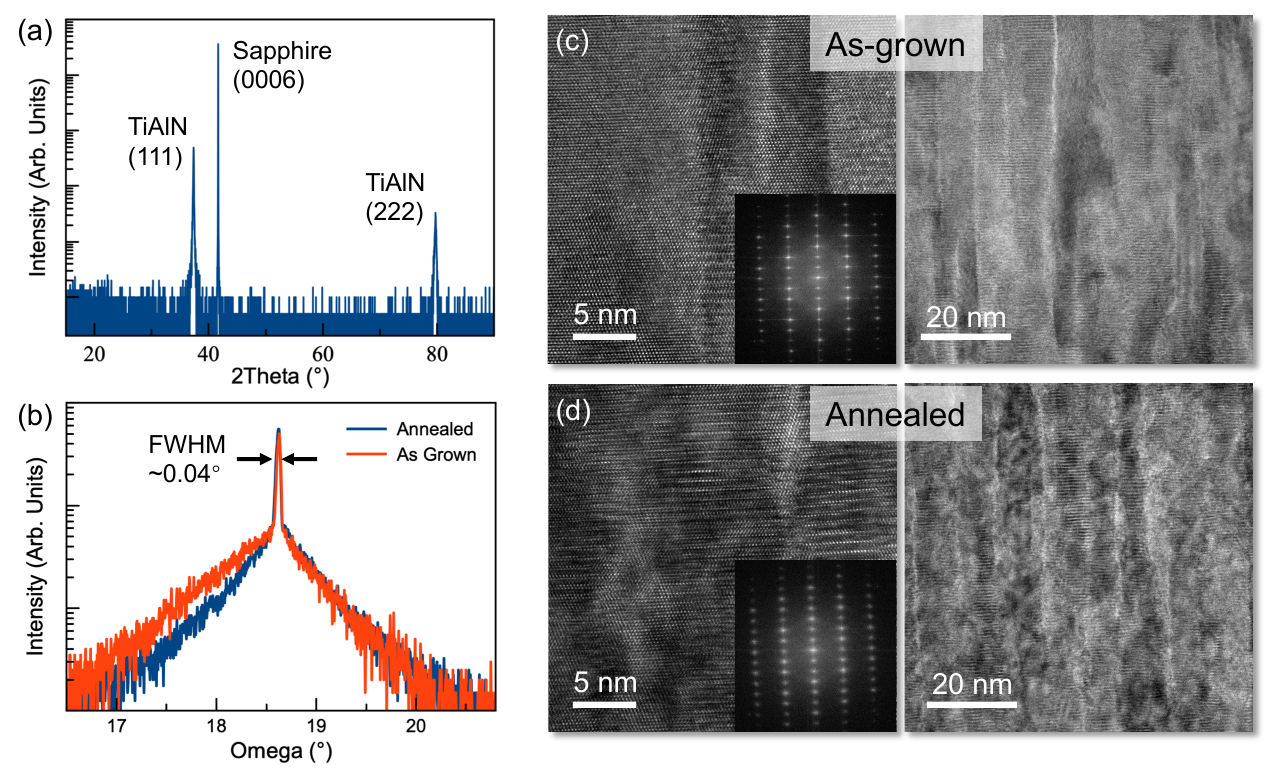}
      \caption{(a) The full-range X-ray line scan on annealed 100 nm-thick TAN films on \textit{c}-cut sapphire substrates. It is clear that no secondary phases were formed and the films maintained epitaxial and good crystallinity. (b) Rocking curves on as-grown and annealed TAN films revealed similar diffraction profiles with a full-width-half-maximum (FWHM) of 0.04 $^\circ$. (c-d) HAADF images of as-grown and annealed TAN films at different magnifications. High-magnification images and fast-\textit{Fourier}-tranform patterns further confirmed the epitaxial nature of the high-quality films. Low-magnification images revealed clear columnar grains for both samples, where the grain dimensions remained essentially unchanged after annealing.}
      \label{fig:TANMaterFigS2} 
\end{figure}

\begin{figure} [h]
      \centering
      \includegraphics[width=12.9cm]{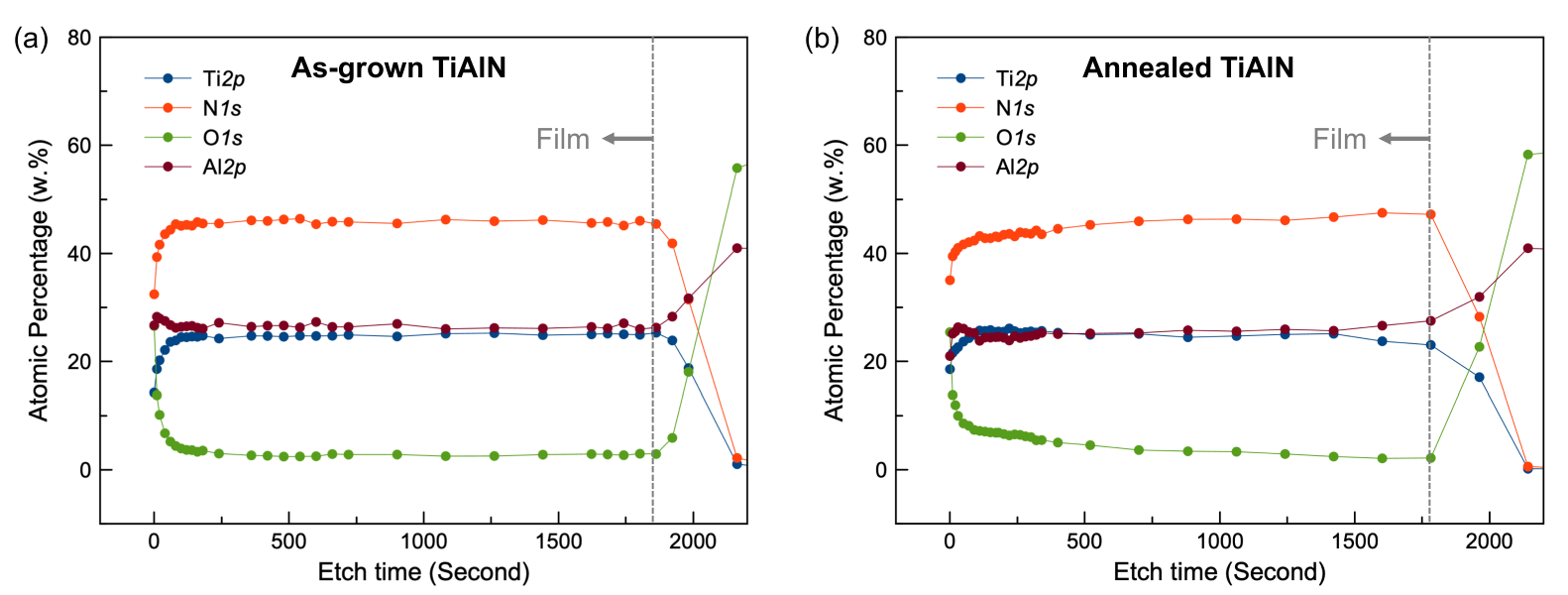}
      \caption{(a) The XPS depth-profile of the as-grown TAN samples. A clear surface oxidation region was observed and the film is largely homogeneous along the out-of-plane direction. (b) The XPS depth-profile of the annealed TAN samples. Again, surface oxides are visible, and an additional gradient of elements was found. More specifically, from the interface to film surface, the concentration of aluminum and nitrogen is gently decreasing while the concentration of titanium and oxygen is gentaly increasing. }
      \label{fig:TANMaterFigS3} 
\end{figure}

Additional characterization data on TAN thin films including X-ray line scans, rocking curves, TEM scans, and XPS depth-profiles scans are provided (Supplementary Fig. \ref{fig:TANMaterFigS2},\ref{fig:TANMaterFigS3}). 

\newpage
\section{Device patterning processes}

\begin{figure} [h]
      \centering
      \includegraphics[width=12.9cm]{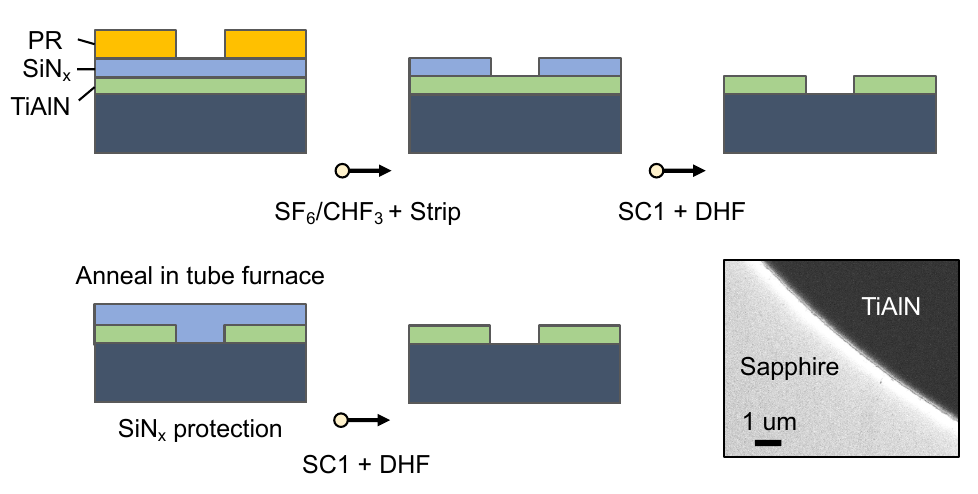}
      \caption{The schematics of the device fabrication process for TAN samples. The SEM image of the device edge profile is given. }
      \label{fig:TANMaterFigS4} 
\end{figure}

The \textit{Hall}-bar devices of TAN were patterned using a developed wet-etch approach, and the microwave resonator devices of TAN were fabricated similar with an additional patterning process on the grounding TiN layers. The detailed description of the process can be found in the main text (Supplementary Fig. \ref{fig:TANMaterFigS4}, see Methods section). Since only SiN\textsubscript{x}-hardmask was applied, which can be easily stripped in the DHF solution, the SEM image on the device edges revealed clean surfaces for both TAN films and the underlying sapphire substrates. 

\newpage
\section{Additional electrical transport data for thick TAN films}

From the \textit{Hall} measurement (see main text, Fig. 2b), it can be inferred that the conduction in annealed TAN is mainly carried by electrons, the density of which is given by
\begin{equation}
    \label{eqn:eqn1}
    \frac{1}{n_e^{10K}ed}=\frac{R_{xy}}{\mu_0H},
\end{equation}
where $n_e^{10K}$ is the electron density at 10 K, $d$ is the film thickness, and $R_{xy}$ is the measured \textit{Hall} resistance as a function of applied field $\mu_0H$. Owing to the high electron density, we then use the free-electron model to extract a set of corresponding material parameters, which can be further used as reference numbers for future studies or comparison with other systems. Specifically, 
\begin{gather}
    k_F=(3\pi^2n_e)^{1/3}, \label{eqn:eqn2}\\
    v_F=\frac{\hbar k_F}{m_e}, \label{eqn:eqn3}\\
    \tau=\frac{m_e}{n_ee^2\rho_{xx}}, \label{eqn:eqn4}\\
    l=v_F\tau, \label{eqn:eqn5}
\end{gather}
where $k_F$ is the wavevector at the \textit{Fermi} surface, $v_F$ is the \textit{Fermi} velocity, $\tau$ is the relaxation time or elastic scattering time, $\rho_{xx}$ is the longitudinal resistivity, and $l$ is the elastic mean free path of electrons. Similar measurements were also performed on a 100 nm-thick TiN films (same deposition conditions, data not shown here) and the extracted parameters were listed for both materiral systems (Table \ref{tab:Table1}). The \textit{Ioffe-Regel} criterion $k_Fl$ is approaching unity for annealed TAN, indicating that the material is a strongly disordered system. It is also worth noting that, the annealed TAN was treated as one entity and the extracted parameters are essentially an average of the microscopic regions. While from a microscopic view, the annealed TAN is expected to have spacial variation in charge-carrier density, which could potentially make this material resembles a percolating superconductor \cite{Deutscher1980,Strelniker2007}.

\begin{figure} [h]
      \centering
      \includegraphics[width=12.9cm]{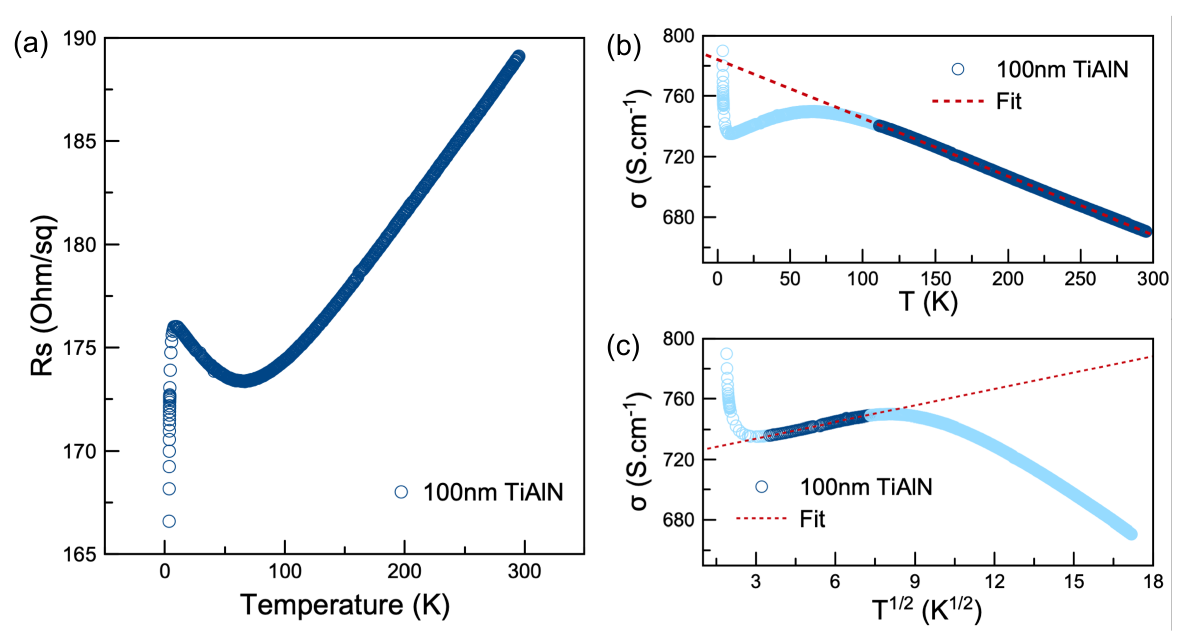}
      \caption{(a) A zoomed $R_s(T)$ curve of annealed 100 nm-thick TAN films. There is a clear "cusp" due to the transition of $dR_s/dT$, a characteristic feature of disordered metals. (b) From 300 K to $\sim100$ K, the conductivity has a linear dependency on temperature. The fitting has a $R^2$ of 0.9998. (c) From $\sim10$ K to $\sim60$ K, the conductivity has a linear dependency on the square root of temperature. The fitting has a $R^2$ of 0.9977.}
      \label{fig:TANMaterFigS5} 
\end{figure}

   \begin{table}[h]
      \caption{\label{tab:Table1}The measured and extracted material parameters of 100 nm-thick annealed TAN and TiN.}
      \begin{ruledtabular}
        \begin{tabular}{cccccccc}
          Materials & $d$ (nm) & $T_c$ (K) & $R_s$ ($\Omega$) & $n_e^{10K}$ ($\times10^{28}m^{-3}$) & $L_k^{\square}$ (pH/$\square$) & $l$ (nm) & $k_Fl$ \\
          \colrule
          TiN & 98 & 3.8 & 8.5 & 4.46 & 3 & 1.33 & 14.6\\
          Ti\textsubscript{0.5}Al\textsubscript{0.5}N & 100 & 3.2 & 132.3 & 1.60 & 57 & 0.18 & 1.38\\
        \end{tabular}
      \end{ruledtabular}
    \end{table}

The nature of a disordered metal can also be inferred from the $R_s(T)$ curves (Fig. \ref{fig:TANMaterFigS5}). At high temperatures (from $\sim100$ K to 300 K), the conductivity has linear dependency on $T$ with a negative $d\sigma/dT$ (Supplementary Fig. 5b), indicating the metallic nature of the material where electron scattering is limited by electron-phonon interactions. While at the intermediate temperature regime ($\sim10$ K to $\sim60$ K), the conductivity has linear dependency on $T^{1/2}$ with a positive $d\sigma/dT$ (Supplementary Fig. 5c), implying that the electron transport is limited by the electron localization \cite{Lee1985,Kaveh1982}. 

\begin{figure} [h]
      \centering
      \includegraphics[width=12.9cm]{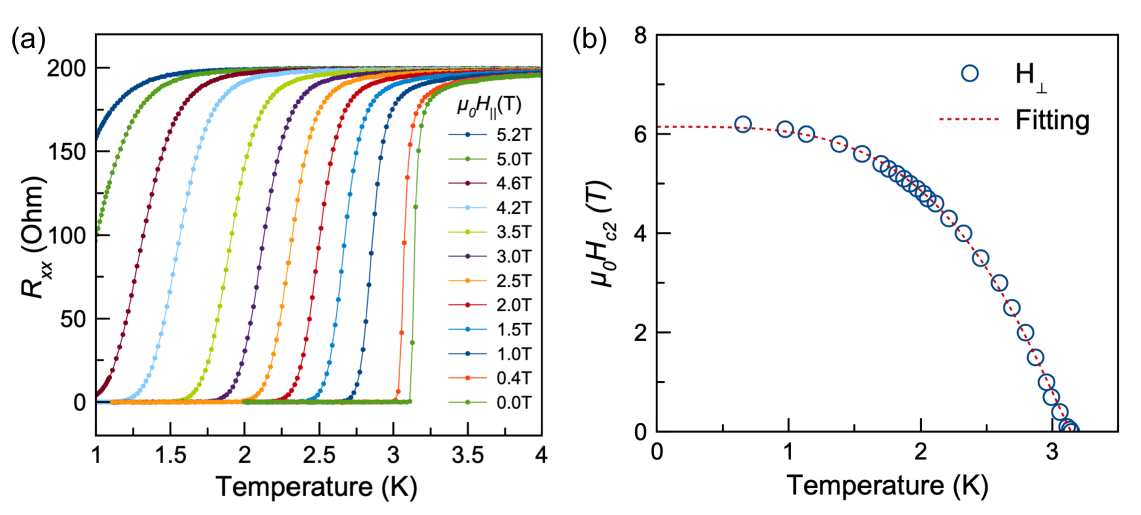}
      \caption{(a) The $R_{xx}(T)$ measured under parallel fields. (b) The relationship between the upper critical ($H_{c2}^{\perp}$) and superconducting transition temperature ($T_c$) was given for perpendicular fields. The dotted line is the fitted curve. }
      \label{fig:TANMaterFigS6} 
\end{figure}

The temperature-dependent longitudinal resistance as a function of applied fields in parallel to the sample surface is provided (Supplementary Fig. \ref{fig:TANMaterFigS6}). 

In addition, the relationship between the upper critical ($H_{c2}^{\perp}$) and superconducting transition temperature ($T_c$) under perpendicular fields deviates from the \textit{GL} fomula, and was fitted using an empirical power-law formula $\mu_0H_{c2}^{\perp}(T)=\mu_0H_{0}^{\perp}[1-(T/T_c)^\alpha]^\beta$, where $\alpha$ = 3.6 and $\beta$ = 1.1. The fitting yields a slight upward curvature at low fields and saturates at high fields. Although the physical significance of this fitting model is yet clear, a detailed discussion on the proposed mechanism was given in the main text.

\clearpage
\section{Crygenic measurement setup}

\begin{figure} [h]
      \centering
      \includegraphics[width=8.6cm]{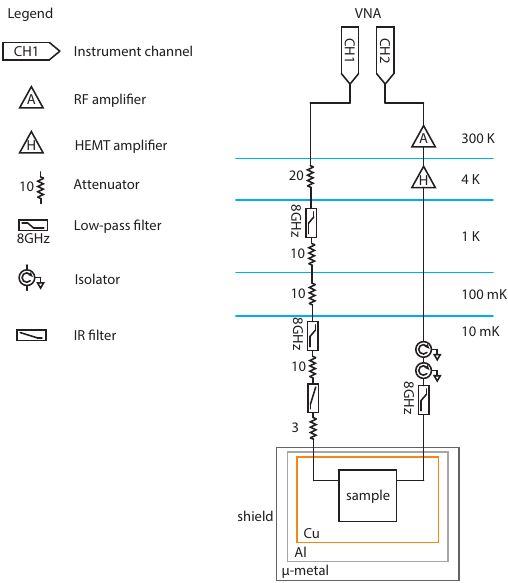}
      \caption{The cryogenic setup for the microwave-resonator measurement.}
      \label{fig:TANMaterFigS7} 
\end{figure}

The cryogenic measurement setup for the microwave-resonator measurement is given (Supplementary Fig. \ref{fig:TANMaterFigS7}). 

\clearpage
\section{Electrical transport data for thin TAN films}
\begin{figure} [h]
      \centering
      \includegraphics[width=8.6cm]{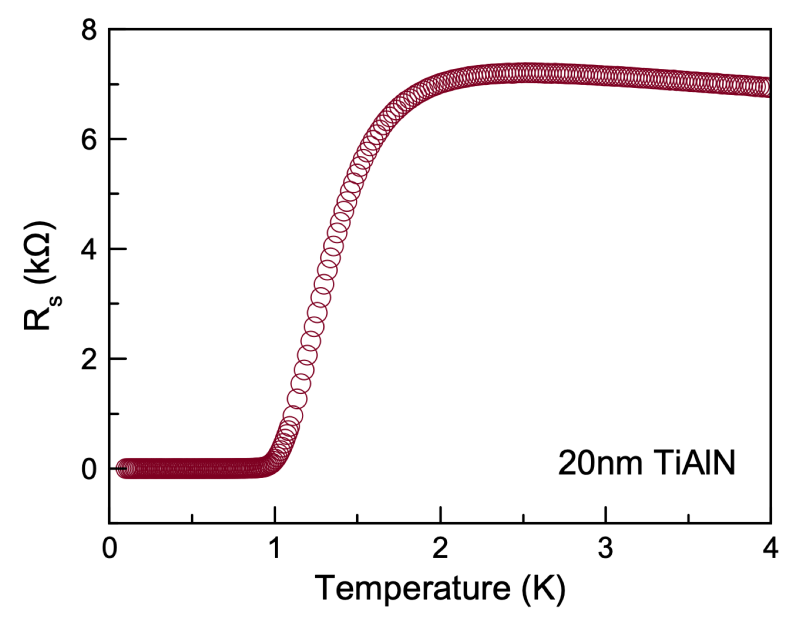}
      \caption{The sheet resistance of 20 nm-thick annealed TAN films near superconducting transition temperature.}
      \label{fig:TANMaterFigS8} 
\end{figure}

20 nm-thick TAN deposited and annealed under the same conditions was patterned and measured (Supplementary Fig. \ref{fig:TANMaterFigS8}). A large sheet resistance was observed, corresponding to a sheet kinetic inductance of $\sim7.5$ nH/$\square$ using the mid-point of the superconducting transition edge as $T_c$. This is essentially a demonstration of the applicability of spinodal TAN for ultrahigh-impedance circuit applications. Detailed characterization on physical properties of thinner TAN is under further investigation.  

\bibliography{TANMater_SuppRef}